\documentclass[titlepage,twoside,12pt]{article}
\usepackage{amssymb}
\usepackage{amsfonts}
\begin{document}

{\bf \Large Cosmic Contact : To Be, or \\ \\ Not To Be Archimedean ?} \\ \\

Elem\'{e}r E Rosinger \\ \\
Department of Mathematics \\
and Applied Mathematics \\
University of Pretoria \\
Pretoria \\
0002 South Africa \\
eerosinger@hotmail.com \\ \\ \\

{\bf Abstract} \\

This is a two part paper which discusses various issues of cosmic contact related to what so
far appears to be a self-imposed censorship implied by the customary acceptance of the
Archimedean assumption on space-time. \\ \\ \\

{\bf \Large Part I : Cosmic Contact Censorship : \\ \\
an Archimedean Fallacy ?} \\ \\

{\bf Abstract} \\

It is argued that the customary, and rather tacitly taken for granted, assumption of the
Archimedean structure of physical space-time may be one of the reasons why we experience
Cosmic Contact Censorship. Further, it is argued that, once a non-Archimedean view of physical
space-time is adopted, a variety of alternative worlds becomes open, a variety which may in
part explain that Cosmic Contact Censorship. \\ \\ \\

{\bf 0. Preliminaries} \\

There is a well known literature on issues such as : "are we alone in the universe ?", "how
many civilizations are out there in the galaxy or beyond ?", and so on, see [4] and its
references for some of the more recent such contributions. \\
For convenience and brevity, and following the implicit suggestion of [4], let us call such
issues CCC. \\

One of the familiar arguments when debating CCC issues is that, quite likely, life and/or
intelligence in Cosmos, if any to exists beyond Planet Earth, need not necessarily be confined
to its forms known to us so far on our planet. And if such may indeed be the case, then quite
obviously we can face a considerably difficult issue, having to search for, and eventually
recognize what we quite likely have no absolutely any idea about. \\

In this paper, however, another limitation in debating CCC issues is addressed, one that, so
far, appears to have been missed altogether. Namely, it is related to what may turn out to be
the excessive limitations in our conditioning as manifested in our usual perceptions and
conceptions of space and/or time. Fortunately, this second limitation can be clarified much
more easily, since it can be formulated in rather simple mathematical terms which, even if
only intuitively, happen to be familiar to all of us. \\

{\bf 1. Walking Inside the Traditional Archimedean Trap} \\

As it happens, rather by an omission or default, than by any more conscious and deliberate
commission, all sides involved in CCC arguments and disputes, whether supporting or denying
the uniqueness in Cosmos of life and/or civilization on Planet Earth, seem to take rather for
granted a Euclidean sort of mathematical model of space-time, and on occasion, its general
relativistic version. \\

Needless to say, there is in fact a strong motivation for such a position, since until the
beginning of the 20th century, science, and in particular mathematics, did not know or care
much about space-time structures which were not Euclidean, or at least, were not locally
so. \\

A remarkable fact in this regard, hardly ever considered according to its possible relevant
implications, is that an essential feature of such type of space-time structures is in their
being {\it Archimedean}. And this feature may turn out to be highly relevant to the issues of
CCC. \\

In the simplest, one dimensional case of a Euclidean space, namely, of the real line
$\mathbb{R}$, the Archimedean property simply means that there exists a positive real number
$u \in \mathbb{R},~ u > 0$, such that for every real number $x \in \mathbb{R}$, there exists a
positive integer $m \in \mathbb{N},~ m \geq 1$, with the property that $ m u \geq |\,
x\,|$. \\
Of course, we can for instance choose $u = 1$, or for that matter, any other strictly positive
$u \in \mathbb{R},~ u >0$. And then the Archimedean property simply means that, no matter
where the point $x$ would be on the real line $\mathbb{R}$, we can in a {\it finite} number of
steps walk past $x$, if we start at the origin $0 \in \mathbb{R}$, and our steps are of length
$u$. \\

Needless to say, geometry, especially as practiced at its historical origins in ancient times,
could only be of practical use if it assumed the Archimedean property for the real line. \\

This Archimedean property extends naturally to Euclidean spaces, that is, to finite dimensional
vector spaces over the real numbers $\mathbb{R}$. Indeed, on $\mathbb{R}^n$, with $n \in
\mathbb{N},~ n \geq 2$, we have the following natural {\it partial order}. Given $x = ( x_1,
\ldots , x_n ),~ y = ( y_1, \ldots , y_n ) \in \mathbb{R}^n$, then \\

$~~~~~~ x ~\leq~ y ~~\Longleftrightarrow~~ x_1 \leq y_1, \ldots , x_n \leq y_n $ \\

Now if we take for instance $u = ( 1, \ldots , 1 ) \in \mathbb{R}^n$, then for every $x =
( x_1, \ldots , x_n ) \in \mathbb{R}^n$, there exists $m \in \mathbb{N},~ m \geq 1$, such that
$m u \geq |\, x \,|$, where $|\, x \,| = ( |\, x_1 \,|, \ldots , |\, x_n \,| )$. \\

In particular, the set $\mathbb{C}$ of complex numbers, which as vector space over
$\mathbb{R}$ is isomorphic with $\mathbb{R}^2$, also enjoys the Archimedean property. \\

Here we should further note that, on top of the practical geometric considerations, there is a
deeper, and purely mathematical reason for us humans having ended up historically with such a
fundamental, and in fact, exclusive role played by the real line $\mathbb{R}$ in mathematics
and physics. Namely, as is well known in algebra, $\mathbb{R}$ is the only linearly ordered
complete field which is Archimedean. \\ \\

{\bf 2. Beyond the Archimedean Conundrum} \\

As mentioned however in [3], and in the literature cited there, recently there has emerged an
interest in physics in considering mathematical models which use {\it other} scalars than the
traditional real or complex numbers. \\
The reasons for such a venture may be numerous and varied. However, several pointers in this
regard can be recognized as rather remarkable in being thought provoking. \\

One of them, of a markedly general and deep nature, is the question posed in [1] and asking
how it comes that, so far, all the spaces used in physics, including general relativity and
quantum theory, have a cardinality not larger than that of the continuum, that is, of the set
$\mathbb{R}$ of real numbers ? \\
After all, ever since Cantor's set theory introduced in the late 1800s, we know about sets with
cardinals incomparably larger than that of the continuum. Not to mention that the cardinal of
the continuum is merely one the smallest infinite cardinals, and in fact, it is but the very
second one, if we accept the Continuum Hypothesis. \\
Yet quite unfortunately so far, no one seems to be able to come forward with a credible answer
to that question ... \\

A second pointer, perhaps somewhat more near to home, yet no less hard to disregard, arose in
the 1960s, with the introduction of Nonstandard Analysis by Abraham Robinson. \\
Motivated by the need to create a rigorous mathematical theory for the "infinitesimals" used
so astutely and effectively by Leibniz in Calculus back in the late 1600s, Robinson
constructed an extension of the real line $\mathbb{R}$. This extension $^*\mathbb{R}$, called
the {\it nonstandard reals}, is a linearly ordered field, just like $\mathbb{R}$ itself,
however, it is - and as follows from the argument mentioned above, must be - {\it
non-Archimedean}. \\

There have, of course, been several other candidates for sets of scalars which were suggested
for use in the mathematical modelling of physics. Some of them are presented in the literature
cited in [3]. \\

The remarkable fact in this regard, and so far often missed, is that there is an arbitrarily
large pool of sets of scalars which could be taken in consideration for the mathematical
modelling of physics. This variety, described in [3], is given by a rather easy and ubiquitous
mathematical construction. As it happens, this construction, in several of its particular
instances, is already known by many in mathematics, without however the widespread enough
realization of the existence of a deeper underlying and unifying method. Indeed, the very
construction of the real numbers from the rational ones, according to the Cauchy-Bolzano
method introduced in the 1800s, is but one such instance, as it is the way metric spaces, or
in general, uniform topological spaces, are completed in modern topology. \\
That deeper underlying unifying method is called "reduced powers" in terms of Model Theory,
which is a branch of Mathematical Logic. \\
By the way, a particular and technically rather involved case of such reduced powers, called
"ultrapowers", can be used in the construction of nonstandard reals $^*\mathbb{R}$ as well. As
for the more general "reduced powers", their construction and use is significantly simpler. \\

In general, the mentioned reduced power construction can lead to {\it algebras} $A$ of {\it
scalars} which, unlike is the case with both $\mathbb{R}$ and $^*\mathbb{R}$, are no longer
fields. In other words, these algebras $A$ have "zero divisors", which means that in such
algebras, and unlike in fields, one can have elements $a,~ b \in A$ whose product is zero,
that is, $a . b = 0$, without $a$ or $b$ being zero. Consequently, in such algebras one cannot
divide with every nonzero element. \\
However, such a restriction is {\it not} strange at all, since the same happens already with
usual matrices. Furthermore, and unlike with matrices, such algebras $A$, if desired, can be
constructed so as to have a commutative multiplication. \\

A remarkable feature of such reduced power algebras $A$ is that they contain "infinitesimal"
type elements, and as a consequence, they also contain elements which are "infinitely large".
This leads to the fact, just like in the case of the nonstandard reals $^*\mathbb{R}$, that
such algebras $A$ are non-Archimedean. \\ \\

{\bf 3. Universes within Universes ... ad infinitum ...} \\

In order to have a somewhat easier understanding of the effects of the non-Archimedean
property as they may relate to the issues of CCC, let us return to the simplest one
dimensional case of the nonstandard reals $^*\mathbb{R}$, and use it as an intuitive mental
model, rather than the more richly structured reduced power algebras. Fortunately however, for
this purpose, we do not have to get involved with the often elaborate technical details of
Nonstandard Analysis, which are quite considerable when compared with the general construction
of reduced power algebras presented in [3]. \\

As noted above, the intuitive essence of the Archimedean property is that, in a finite number
of steps, one can walk past every point in the respective space, no matter where one started
to walk. In this way, an Archimedean space, like for instance the real line $\mathbb{R}$, is
but {\it one single world}. \\

On the other hand, in a non-Archimedean space, such as that of the nonstandard reals
$^*\mathbb{R}$, or of the reduced power algebras, one is {\it inevitably confined} to a very
small part of that space when walking any finite number of steps, with the steps no matter how
large, but of a given length. It follows that non-Archimedean spaces, among them the reduced
power algebras, contain {\it many worlds} which are inaccessible to one another by the
mentioned kind of walking, or at best, one of them is accessible to the other, but only in a
most limited manner. \\

Let us try to clarify somewhat more this issue of accessibility, without however getting
involved here in technical complications. For convenience, we denote by $WW_{u,\,x}$ the part
of the non-Archimedean space, be it $^*\mathbb{R}$ or a reduced power algebra $A$, which can
be accessed from a given point $x$ through walking a finite number of steps of size $u >
0$. \\
Of course, it is easy to see that if we take any point $y \in WW_{u,\,x}$, then \\

$~~~~~~ WW_{u,\,y} ~=~ WW_{u,\,x} $ \\

Further, it follows easily that the nonstandard reals $^*\mathbb{R}$ and the reduced power
algebras $A$ do in fact contain {\it infinitely many} disjoint "walkable worlds" $WW_{u,\,x}$,
each two of them being inaccessible to one another. \\

And as if to add to the surprises and wonders of such non-Archimedean spaces, such worlds
$WW_{u,\,x}$ can not only be outside of one another, but they can be {\it nested} within one
another in infinitely long chains. This is but a simple and direct algebraic effect of the
fact that some "infinitesimals" can be infinitely larger, or for that matter, infinitely
smaller, than other "infinitesimals". Similarly, some "infinitely large" elements can be
infinitely smaller, or alternatively, infinitely larger than other "infinitely large"
elements. \\

For instance, in the case of the nonstandard reals $^*\mathbb{R}$, let us take $u = 1,~ x = 0
\in \mathbb{R} \subsetneqq\, ^*\mathbb{R}$. Then $WW_{u,\,x} \supsetneqq \mathbb{R}$, yet it
is known that $WW_{u,\,x}$ is but a tiny part of the whole of $^*\mathbb{R}$. In fact, if we
take $v \in\, ^*\mathbb{R} \setminus \mathbb{R}$, then again $WW_{u,\,x} \subsetneqq\
WW_{v,\,x}$, with the former being but a tiny part of the latter. Furthermore, the latter is
still a tiny part of the whole of $^*\mathbb{R}$. \\
And as it happens, each of the walkable worlds $WW_{u,\,x}$, no matter how one would choose
$u,~ x \in\, ^*\mathbb{R}$, is but a tiny part of the whole of $^*\mathbb{R}$. \\

Added to this comes the story of infinitesimal walkable worlds. For instance, if we take $u =
1,~ v = \epsilon,~ x = 0 \in\, ^*\mathbb{R}$, where $\epsilon > 0$ is a nonstandard positive
infinitesimal, then $WW_{v,\,x} \subsetneqq WW_{u,\,x}$, and the former is again only a tiny
part of the latter. \\
However, we can take both $u,~ v > 0$ to be positive infinitesimal, and we can further assume
that $v / u$ is itself an infinitesimal. In that case we shall again have $WW_{v,\,x}
\subsetneqq WW_{u,\,x}$, with the former once more but only a tiny part of the latter. \\

In the case of the nonstandard reals $^*\mathbb{R}$, we can conclude as follows. Given $u,~
v,~ x,~ y \in\, ^*\mathbb{R},~ u,~ v >0$, the corresponding walkable worlds $WW_{u,\,x},~
WW_{v,\,y}$ can be in one and only one of the next three situations : \\

1)~~~ $ WW_{u,\,x} ~=~ WW_{v,\,y} $ \\

2)~~~ $ WW_{u,\,x} \bigcap WW_{v,\,y} ~=~ \phi $ \\

3)~~~ $ WW_{u,\,x} \bigcap WW_{v,\,y} ~\neq~ \phi,~~~ WW_{u,\,x} ~\neq~ WW_{v,\,y} $ \\

in which case, either \\

~~~~~~3.1)~~~ $ WW_{u,\,x} ~~\mbox{is an infinitesimal part of}~~ WW_{v,\,y} $ \\

or \\

~~~~~~3.2)~~~ $ WW_{v,\,y} ~~\mbox{is an infinitesimal part of}~~ WW_{u,\,x} $ \\

Furthermore, the situation at 2) does happen infinitely many times, and in particular, each
walkable world  $WW_{u,\,x}$ is merely an infinitesimal part of the whole of $^*\mathbb{R}$.
As for the situation at 3), the respective nestings of walkable worlds have infinite length. \\

Needless to say, in the case of reduced power algebras which, as mentioned, have a richer
structure than the nonstandard reals $^*\mathbb{R}$, the above three situation manifest
themselves in yet more complex ways. \\ \\

{\bf 4. Do We Live in One Universe ? Are the Quanta the \\
\hspace*{0.45cm} Smallest Possible Entities ? And what about Time ?} \\

It is quite remarkable, although often missed to be noted, or in fact simply disregarded, that
much of classical mechanics is subjected to what is called Dimensional Analysis, [2]. In other
words, all respective physical entities can be defined in terms of only three fundamental ones,
namely, {\it length, mass} and {\it time}. Furthermore, in terms of the respective definitions,
all the corresponding physical entities are elements of {\it scaling groups}, which means that
there is {\it no} natural, unique or canonical way to choose their units, and on the contrary,
those units can be chosen arbitrarily, and merely upon convenience. \\

This clearly implies that each of the three fundamental physical entities is supposed to belong
to an Archimedean space, namely, $\mathbb{R}$ in the case of length and time, and $[ 0, \infty
) \subsetneqq \mathbb{R}$, in the case of mass. \\

As for quantum mechanics, such an approach is of course no longer accepted, due to the
radically different assumption of the existence of {\it minimal} values for various physical
entities involved, values called the respective "quanta". \\

And yet, the passing from classical mechanics to quantum mechanics has not led to the
abandonment of the Archimedean assumption. And the fact is that, as things stand so far, it
did not have to do so. Indeed, all what happened was that in the case of quantized physical
entities, corresponding intervals of real numbers were simply excluded from the real line
$\mathbb{R}$. For instance, let $q > 0$ be the quantum quantity for a certain physical entity,
then instead of the respective quantity being able freely to range over the whole of
$\mathbb{R}$ as it may happen in classical mechanics, now it is only allowed to do so over the
{\it discrete} subset of $\mathbb{R}$, given by $\mathbb{Z} q$, that is, the integer multiples
of $q$. \\

Given such a state of affairs, including in general relativity and quantum mechanics, it is no
wonder that in cosmology we still assume, even if not explicitly and up front, that real,
physical space is exhausted by $\mathbb{R}^3$, or rather, by some curved general relativistic
version of it, while real, physical time is like $\mathbb{R}$. \\
In other words, we still think within the limitations of a {\it one world} Archimedean world
view ... \\

And then the question arises :

\begin{itemize}

\item What if indeed we may in fact live in non-Archimedean worlds, be it space-wise, or
time-wise, or for that matter, in both of these ways ?

\end{itemize}

And if it may happen that we do live in such non-Archimedean worlds, then the respective
alternatives 1) - 3) in section 3 may actually apply. Not to mention that in case reduced
power algebras more rich in structure than the nonstandard reals may be adequate for modelling
physics, yet more complex alternatives could be encountered. \\

And quite clearly, the mentioned alternative 2) already bring in a dramatic situation related
to CCC. Indeed, it is hard to imagine what kind of communication may ever take place between
two such disjoint walkable worlds, be they disjoint space-wise, time-wise, or in both of these
ways ... \\
Interestingly enough, the situation is not much simpler in the case of alternative 3), that is,
even if two walkable worlds may happen to have a common part. Namely, in such a case one of
such worlds must be contained in the other, but then, it is contained as a mere infinitesimal
part. Therefore, again, it is hard to imagine what kind of communication may ever take place
between two such walkable worlds ... \\

Finally, let us note that, especially related to {\it time} in the above alternative 3) there
seem to be immense difficulties with respect to CCC between such two walkable worlds $WW$ and
$WW\,'$. \\
Namely, if we are in the walkable world $WW$ which infinitesimally small compared with $WW\,'$,
then during our own time quite nothing seems to happen in $WW\,'$, due to the respective
infinite disproportion between the time scales involved. In this way, we in $WW$ may see
$WW\,'$ as merely frozen, dead, or immobile ... \\
Conversely, if $WW\,'$ is infinitesimally small compared with our walkable world $WW$, then
events in $WW\,'$ may happen infinitely fast when seen in our own time scales. Therefore,
again, we may simply not be able to take notice of them, thus once more seeing $WW\,'$ as
merely frozen, dead, or immobile, even if because of the totally opposite reason ... \\

The fact is that, those among us who have for a while been working in Nonstandard Analysis, or
with reduced power algebras, do not feel anything strange about the kind of rather complex
compartmentalization of walkable worlds described in 1) - 3) in section 3. \\

And needless to say, Nonstandard Analysis has during the last four decades proved its
remarkable value both in mathematics and its applications, among the latter, stochastic
analysis, [5]. \\

As for various reduced power algebras, beyond the scalar ones used in [3], they have over the
last more than four decades proved their utility in solving very large classes of earlier
unsolved linear and nonlinear PDEs, [6-22]. Indeed, such reduced power algebras can give an
infinitely large class of differential algebras of generalized functions, each containing the
Schwartz distributions. The respective differential algebras of generalized functions, and
among them in particular, the so called Colombeau algebras, proved to be able to provide for
the first time in the literature suitable generalized solutions within a systematic {\it
nonlinear} theory of generalized functions, a theory not available within the classical
Sobolev or Schwartz linear distribution theories.

\newpage

{\bf \Large Part II : On Cosmic Contact Self-Censorship} \\ \\

{\bf Abstract} \\

Before delving into the issue of cosmic contact, or its possible censorship due to various
sources, it is important to clarify as much as possible the meaning of the concept of such,
or for that matter, any other relevant possible contact. Without a more appropriate a priori
clarification, it is most likely that we ourselves may actually enforce a censorship, even
if we do so not consciously. This paper points to several possible conceptual obstacles in
the venture of clarifying as much as possible the meaning of contact, be it cosmic or of
other nature. Such a clarification is seen as a necessary step in order to avoid unintended
self-censorship. In particular, in case we may at last consider non-Archimedean space-time
structures as well, then what we usually call "Cosmos" may in fact happen to be everywhere
inside and nearby, all around us, as well as at distances never imagined in our usual
Archimedean paradigms. This, in its remarkable richness and complexity, is in stark
contradistinction with the poverty of "one single Cosmos, and out there" typical of the
Archimedean vision. \\ \\

{\bf 1. Preliminaries} \\

Cosmic contact, as in fact any sort of contact, can have a large variety of meanings. And by
missing to be aware of specific possible meanings, we significantly increase the likelihood of
exerting a de facto, even if not conscious, self-censorship. \\
In this regard, there can at least be two ways in which our meaning of contact suffers from
restrictions. One of them, quite likely by far the most difficult to overcome, is the overall
limitation of human awareness at any specific given time. The other one, possibly easier to
deal with, is due to the limitations we impose, and do so without being conscious about that
fact, upon the assumptions which happen to constitute the conceptual background within which
we are looking for possible meanings for the phenomenon of contact. \\

Related to that second way, the way in which our background conceptual assumptions can limit
the meanings we associate with the phenomenon of contact, it was pointed out in [4] that the
usual Archimedean assumption on the structure of space-time, so prevalent, if not in fact the
only one, in modern Physics, may actually be the source of a major self-censorship, one which
we keep failing to become conscious of. Further details related to this argument were
presented in [2,3]. \\

That second way, which can be the source of much - and at the same time, less than conscious -
self-censorship, has at least two manifestations, namely, in :

\begin{itemize}

\item our background assumptions about "where in space and time we are supposed to be
looking for possible contact", assumptions at present of a near exclusive Archimedean nature,

\end{itemize}

and rather independently of that

\begin{itemize}

\item "what kind of contact" we keep thinking about, thus by implication excluding other
possible variants of it.

\end{itemize}

In [4], the first of these two manifestations was considered, and the lack of a sufficient
awareness about the possibility of a non-Archimedean space-time structure was pointed out,
indicating at the same time the surprising richness and complexity of the {\it self-similar}
nature of non-Archimedean structures. \\
Here we shall consider the second above alternative, and we shall point out that the concept
of contact, thus its meaning as well, can have at least two rather different variants,
namely :

\begin{itemize}

\item direct contact,

\item indirect contact.

\end{itemize}

Furthermore, the second variant can also have at least two significantly different
sub-variants, namely :

\begin{itemize}

\item contact in which the contactees are aware of it,

\item contact in which at least one of the contactees does not become aware of it.

\end{itemize}

{\bf 2. Recalling Briefly a Few Relevant Features of non-Archimedean \\
        \hspace*{0.4cm} Space-Time Structures} \\

The radically more rich and complex features of non-Archimedean space-time structures,
reflected already in the simplest one dimensional case of the nonstandard real line
$^*\mathbb{R}$, are manifested in the corresponding {\it self-similar} structures which recall
essential properties of fractals. This fact, therefore, should already affect our perceptions
and conceptions of {\it time}. When it comes to {\it space}, needless to say, higher
dimensional instances of non-Archimedean structures may become involved, with their yet more
rich and complex features. \\

As for what may appear as the simpler, one dimensional case of time, two of the essential
novelties in its non-Archimedean instances are the following :

\begin{itemize}

\item there are plenty of "times beyond, of before all time", and

\item there are plenty of "times within every single instant of time".

\end{itemize}

Therefore, even if we keep to our Archimedean perceptions and conceptions of space, and only
let in non-Archimedean structures in the one dimensional case of time, we already have a major
problem in establishing the meaning of contact. Indeed, in such a case, entities "beyond or
before time" may be in direct contact with us, yet we may never become aware of that, if we
keep to our present Archimedean background assumption about time. A similar situation can, of
course, happen with entities which exists in "times within every single instant of
time". \\
In particular, a mere usual instant can prove to be nothing short of "eternity" for certain
worlds. And dually in a way, what is "eternity" in our Archimedean perception and conception
of time may be no more than a mere instant, when considered in non-Archimedean contexts. \\

Needless to say, in case space is allowed a non-Archimedean structure, we immediately end up
with far more rich and complex structures in which :

\begin{itemize}

\item there are plenty of "spaces beyond all space", and

\item there are plenty of "spaces within each and every single space point".

\end{itemize}

And such non-Archimedean structures can similarly, if not even more, affect the meaning of
contact. After all, we, in our finiteness in space, as seen from the Archimedean point of view,
can in fact be hosts to infinitely many worlds, worlds which appear to us, and are conceived
by us a mere "negligible infinitesimal" ones. And in a sort of duality, what is the
Archimedean Cosmos for us may in fact be altogether but a "negligible infinitesimal"
realm ... \\

Clearly, what has so far been conceived as cosmic contact, for instance, by projects such as
SETI, is supposed to take place exclusively within an Archimedean space-time structure. And
the contact is only supposed to be between us humans, and on the other hand, entities somewhere
far out there in the Cosmos, or at least, outside of our Planet Earth, but most certainly not
beyond the confines upon time and space the Archimedean assumption imposes. And of course, such
contacts are even less supposed to be with entities within the infinitesimal realms
non-Archimedean space-time structures allow in such an abundance. \\ \\

{\bf 3. One Reason To Be Careful when Deciding what \\
        \hspace*{0.4cm} Space-Time May Really Be} \\

As argued in [5], see also [2], present day Theoretical Physics does so strangely and
systematically disregard, or even worse, what a Descartes used to call "res cogitans". And
such an attitude manifestly flies in the face of most simple phenomena which can be formulated
in rather clear questions. Questions which Theoretical Physics continually fails to consider,
let alone, deal with. Here are some of them, as cited form [5,2]. \\

{\bf 3.1. Within Newtonian Mechanics.} Instant action at arbitrary distance, such as in the
case of gravitation, is one of the basic assumptions of Newtonian Mechanics. This does not
appear to conflict with the fact that we can think instantly and simultaneously about
phenomena no matter how far apart from one another in space and/or in time. However, absolute
space is also a basic assumption of Newtonian mechanics. And it is supposed to contain
absolutely everything that may exist in Creation, be it in the past, present or future.
Consequently, it is supposed to contain, among others, the physical body of the thinking
scientist as well. Yet it is not equally clear whether it also contains scientific thinking
itself which, traditionally, is assumed to be totally outside and independent of all phenomena
under its consideration, therefore in particular, totally outside and independent of the
Newtonian absolute space, and perhaps also of absolute time. \\

And then the question arises : where and how does such a scientific thinking take place or
happen ? \\

{\bf 3.2. Within Einstein's Mechanics.} In Special and General Relativity a basic assumption
is that there cannot be any propagation of action faster than light. Yet just like in the case
we happen to think in terms of Newtonian Mechanics, our thinking in terms of Einstein's
Mechanics can again instantly and simultaneously be about phenomena no matter how far apart
from one another in space and/or time. \\

Consequently, the question arises : given the mentioned relativistic limitation, how and where
does such a thinking happen ? \\

{\bf 3.3. With Quantum Mechanics.} Let us consider the classical EPR entanglement phenomenon,
and for simplicity, do so in terms of quantum computation. For that purpose it suffices to
consider double qubits, that is, elements of $\mathbb{C}^2 \bigotimes \mathbb{C}^2$, such as
for instance the pair \\

$~~~~~~~~~~~~~~~ |~ \omega_{00} > ~=~ |~ 0, 0 > \,+\, |~ 1, 1 > ~=~ $ \\
(3.1) \\
$~~~~~~~~~~~~~~~ ~=~ |~ 0 > \bigotimes |~ 0 > \,+\, |~ 1 > \bigotimes |~ 1 > \,\,\in
                                            \mathbb{C}^2 \bigotimes \mathbb{C}^2 $ \\

which is well known to be entangled, in other words, $|~ \omega_{00} >$ is not of the form \\

$~~~~~~ ( \alpha |~ 0 > + \beta |~ 1 > ) \bigotimes( \gamma |~ 0 > + \delta |~ 1 > ) \in
        \mathbb{C}^2 \bigotimes \mathbb{C}^2 $ \\

for any $\alpha, \beta, \gamma, \delta \in \mathbb{C}^2$. \\

Here we can turn to the usual and rather picturesque description used in Quantum Computation,
where two fictitious personages, Alice and Bob, are supposed to exchange information, be it of
classical or quantum type. Alice and Bob can each take their respective qubit from the
entangled pair of qubits $|~ \omega_{00} >$, and then go away with their respective part no
matter how far apart from one another. And the two qubits thus separated in space will remain
entangled, unless of course one or both of them get involved in further classical or quantum
interactions. For clarity, however, we should note that the single qubits which Alice and Bob
take away with them from the pair $|~ \omega_{00} >$ are neither one of the terms $|~ 0, 0 >$
or $|~ 1, 1 >$ above, since both these are themselves already pairs of qubits, thus they
cannot be taken away as mere single qubits, either by Alice, or by Bob. Consequently, the
single qubits which Alice and Bob take away with them cannot be described in any other form,
except that which is implicit in (3.1). \\
Now, after that short detour into the language of Quantum Computation, we can note that,
according to Quantum Mechanics, the entanglement in the double qubit $|~ \omega_{00} >$
implies that the states of the two qubits which compose it are correlated, no matter how far
from one another Alice and Bob would be with them. Consequently, knowing the state of one of
these two qubits can give information about the state of the other qubit. On the other hand,
in view of General, or even Special Relativity, such a knowledge, say by Alice, cannot be
communicated to Bob faster than the velocity of light. \\
And yet, anybody who is familiar enough with Quantum Mechanics, can instantly know and
understand all of that, no matter how far away from one another Alice and Bob may be with
their respective single but entangled qubits. \\

So that, again, the question arises : how and where does such a thinking happen ? \\

And one quite clear answer to all such questions is that :

\begin{itemize}

\item As far as the Archimedean perception and conception of space is concerned, such thinking
does not much seem to happen or take place anywhere at all ...

\end{itemize}

And in case, it does not in fact happen "outside of space", then quite certainly, it must
happen "outside of time", or at least, outside of the usual Archimedean perception and
conception of it. After all, as mentioned above, Relativity alone would simply not allow it to
happen anywhere in space-time ... \\
Thus, quite likely, we are back to some variant of the Cartesian "res cogitans" ... \\
No wonder, therefore, that modern Theoretical Physics does its best to avoid such
issues ... \\ \\

{\bf 4. Possible Varieties of Indirect Contact ...} \\

Let us start with what may appear as the simplest situation, namely, when two entities $A$ and
$B$ are in contact with a third entity $X$ in the following manner : $X$ is aware of both $A$
and $B$, but neither $A$, nor $B$ is aware of the other two. \\
This situation may nevertheless constitute a certain indirect contact between $A$ and $B$,
since $X$ may in some ways affect $B$, ways depending in part on $A$, and similarly, $X$ may
affect $A$ in ways depending in part on $B$. \\

An obvious, and rather unsettling, feature of such an instance of indirect contact is that the
two entities $A$ and $B$ may be involved in it without ever realizing it. In particular, in
case $X$ happens to be a suitable enough realm for such a possibility, it may easily turn out
that $A$ produces some, so to say, resonances in $X$ which affect $B$ to some extent, and/or a
similar effect may propagate from $B$ to $A$. \\

A remarkable feature of such a kind of indirect contact between $A$ and $B$ is that nearly all
the requirements for the respective contact are on the third party $X$, rather than on the two
assumed contactees $A$ and $B$. \\

The practical implication for us, terrestrial beings on Planet Earth, of the above kind of
indirect contact is that, in fact, we may have for ages by now been involved in certain instances
of it without any awareness about it, and of course, we may continue to do so in the future ... \\

And as far as non-Archimedean space-time structures are concerned, such an indirect contact could
possibly happen between $A$ and $B$ when, for instance, in their own terms, they belong to two
walking worlds where one is infinitesimal with respect to the other, or the two are removed from
one another by an infinitely large distance. \\

However, such indirect contact can easily happen even when $A$ and $B$ are in the same walkable
world but they are not aware of one another, while on the other hand, the third party $X$ is
observing both of them, without $A$ or $B$ becoming aware of that.

Needless to say, the above minimal conditions on $A$ and $B$ for an indirect contact between them
is to a certain extent natural. Indeed, a more direct contact may require suitable qualifications
from $A$ and/or $B$. Consequently, there may be a considerable variety of less indirect and/or more
direct kind of contacts between $A$ and $B$, and such contacts - with or without the involvement
of third parties $X$ -  may be the subject of subsequent studies.

\end{document}